\documentclass[aps, prb, twocolumn,amsmath,amssymb,floatfix, reprint]{revtex4-2}
\usepackage{graphicx}
\usepackage{soul}
\usepackage{svg}
\usepackage{times}
\usepackage{dcolumn}
\usepackage{hyperref}
\usepackage{stackengine}
\usepackage{physics}
\usepackage{comment}
\hypersetup{colorlinks, allcolors=blue}
\renewcommand{\k}{\boldsymbol{k}}

\newcommand{\vecj}{\boldsymbol{j}}

\newcommand{\vecM}{\boldsymbol{M}}

\newcommand{\vecn}{\boldsymbol{n}}

\begin{document}
\title{Orbital splitter effect and spatial resolution of current-induced orbital accumulation}
\author{Niels Henrik Aase}
\thanks{These authors contributed equally to this work}
\email{niels.h.aase@ntnu.no}
\author{Erik Wegner Hodt}
\thanks{These authors contributed equally to this work}
\email{erik.w.hodt@ntnu.no}
\author{Karl Bergson Hallberg}
\author{Asle Sudb{\o}}
\author{Jacob Linder}
\email[Corresponding author: ]{jacob.linder@ntnu.no}
\affiliation{\mbox{Center for Quantum Spintronics, Department of Physics, Norwegian University of Science and Technology, NO-7491 Trondheim, Norway}}

\begin{abstract}
The emergence of an orbital angular momentum (OAM) response to a charge current holds promise for technological applications, allowing electrical control of magnetization dynamics.
Often, the OAM current is invoked in explaining experimental results for very large orbital transport effects, but this is conceptually challenging as the OAM current is not a conserved quantity.
Instead of utilizing the orbital conductivity associated with the non-conserved OAM current, we here use non-equilibrium Green’s functions to directly image the OAM density in real space under an applied electric current bias.
We find strong spatial variations in OAM density, with the lattice acting as a source and sink of OAM.
Moreover, we show that the OAM response depends sensitively on the angle between the charge current and the crystal axis.
This enables the generation of a transverse OAM response in one current direction and solely a longitudinal response in another.
We refer to this as an orbital splitter effect, analogous to the spin splitter effect in altermagnets.
\end{abstract}

\maketitle

\section{Introduction}
The field of orbitronics \cite{Bernevig2005, Go2021} has flourished over the past half-decade.
Its key degree of freedom is the orbital angular momentum (OAM) of electrons, which can be used to carry information and exert control over magnetization dynamics \cite{Jo2024}.
The OAM of electrons play a pivotal, albeit passive, role in spintronics, as it facilitates the relativistic spin-orbit coupling (SOC), giving rise to many SOC phenomena such as the spin-orbit torque \cite{Miron2011, Liu2012}. 
In these effects, the electron spin was long viewed as more important than its orbital degree of freedom, largely due to OAM being quenched in equilibrium.

The shackles of orbital quenching can be shed in multiple ways. 
In non-centrosymmetric systems, a nonrelativistic orbital Rashba coupling can emerge \cite{Park2011, Kim2013}, allowing for a momentum-dependent OAM texture. 
However, time-reversal symmetry prohibits the generation of net OAM.
This can be circumvented by applying an electric field, generating a dissipative current, breaking time-reversal symmetry.
The combination of orbital Rashba coupling and an electric field generates a finite OAM, an effect known as the orbital Edelstein effect \cite{Salemi2019, Johansson2024}, named after its spin counterpart.
However, even with inversion symmetry, driving the system out of equilibrium can be sufficient to induce an OAM response \cite{Go2018}.

Specifically, Ref.\ \cite{Go2018} demonstrated that an orbital Hall effect (OHE) can emerge from the orbital texture, also in centrosymmetric systems.
The analogous effect in spin systems, namely the spin Hall effect (SHE) \cite{Hirsch1999, Sinova2004, Sinova2015}, can also be understood in terms of the spin texture.
However, unlike the OHE, the intrinsic SHE relies on SOC and is therefore precluded in centrosymmetric systems.
Thus, with the OHE, one is not limited to heavy metals, enabling use of materials that are more beneficial from both a device \cite{Atencia2024} and an environmental perspective \cite{Wang2024}.
Furthermore, because the OHE scales with nonrelativistic quantities, it is often estimated to be much larger than its spin counterpart \cite{Kontani2009, Jo2018, Sahu2024}.
The OHE was recently observed in Ti \cite{Choi2023a} and Cr \cite{Lyalin2023} using the magneto-optical Kerr effect (MOKE). 
The orbital Hall effect and orbital magnetization induced by an electric field have also been studied \cite{cysne_prb_21, cysne_prl_21,  cysne_prb_22,  cysne_prb_23, costa_prl_23} in transition metal dichalcogenides.
As MOKE measures the surface magnetization, these studies offered direct evidence for the OHE, with the caveat that SOC is weak in both materials, making OAM the dominant contributor to the magnetic signal.

An indirect consequence of the OHE is the orbital torque \cite{Go2020}, which fulfills the promise of orbitronics regarding the control of magnetization using electric currents.
The orbital torque has been observed in a range of different systems \cite{Lee2021, Hayashi2023a, Bose2023, Liu2023, Moriya2024}, but it is generally difficult to disentangle from the spin-orbit torque, especially since orbital-to-spin conversion is believed to significantly boost the orbital torque \cite{Ding2020, Go2020}.
This conversion makes multilayer systems an ideal platform for maximizing the orbital torque efficiency, where an orbital current is generated in one layer through the OHE and subsequently converted in an adjacent layer through SOC.
The size and efficiency of the orbital torque make orbital-assisted spin-orbit torque a promising candidate for next-generation memory technology \cite{Gupta2025}.

Still, fully harnessing the potential of the electron OAM necessitates a more comprehensive theoretical understanding of its transport properties. 
One of the most pressing challenges in this is the OAM current, which is in general not a conserved quantity \cite{Haney2010, Go2020a, Han2022, Urazhdin2023, Atencia2024a}.
Recently, a \textit{proper} OAM current was introduced \cite{Go2024}, inspired by the proper spin current \cite{Shi2006}.
It aims to rectify the non-conservation of the OAM, ensuring a reciprocal relation between the proper OAM current and charge current, although the reciprocity may be violated locally \cite{Go2024}.
However, the conventional definition of OAM current remains non-conserved, as there can be an exchange of OAM between the OAM current and the lattice \cite{Go2020a, Han2022, Aase2024}.

Since the OAM current underpins both the OHE and the role of OAM in assisting spin-orbit torque, it is of interest to understand how the non-conservation of the conventional OAM current affects the OAM accumulation. 
Moreover, the orbital conductivity is often used as a figure of merit for quantifying the OHE.
However, because it is associated with the non-conserved OAM current, it remains unclear to what extent a nonzero orbital conductivity will result in OAM accumulation, as this will depend on the orbital relaxation length.
Theoretical studies have estimated this length to be on the order of a few lattice sites in some systems \cite{Urazhdin2023, Rang2024}, suggesting that accumulation effects may be small.
Conversely, experimental results for long-range orbital effects have been interpreted as indications that the OAM currents do persist over long distances \cite{Hayashi2023a}.
Therefore, considering the local OAM density induced by a charge current possibly offers a different perspective than the orbital conductivity.

In our work, we elucidate this by employing a non-equilibrium tight-binding framework, which allows us to directly image, in real space, the emergence of local intra-atomic OAM density under an applied voltage bias.
In this way, we can compare the OAM response to the predictions of the orbital conductivity.
Moreover, this response is directly measurable in experiments, unlike the OAM current. 
With the flexibility our framework offers, we also identify the orbital splitter effect, where the OAM response depends on the angle between the charge current and the crystal axis.

The paper is structured as follows: In Sec.\ \ref{sec:orb_texture} the multiorbital system we consider is introduced, and we discuss how its orbital texture may manifest in terms of OAM transport.
In Sec.\ \ref{sec:non_eq} the non-equilbrium framework is presented.
Applying this framework to our model, we present and discuss our results in Sec.\ \ref{sec:results}, and finally, we summarize our work in Sec.\ \ref{sec:conclusions}.

\section{System and orbital texture}\label{sec:orb_texture}
We employ a multiorbital tight-binding model and consider $d$ orbitals belonging to the $t_{2g}$ sector, namely $d_{yz}$, $d_{xz}$, and $d_{xy}$ orbitals, on a square lattice.
Associated with each orbital $\alpha$ and lattice site $i$ are the creation and annihilation operators $c_{i\alpha}^\dagger$ and $c_{i\alpha}$, such that that a natural basis is $\tilde{c}_i=\begin{pmatrix}
    c_{i,yz} & c_{i,xz} & c_{i,xy}
\end{pmatrix}^{\mathrm{T}}$
. While these orbitals are, in their own right, $L=2$ orbitals, together they form an effective $L=1$ subspace \cite{Mercaldo2020}.
Without SOC, both $L$ and spin $S$ remain good and separate quantum numbers.
As we focus on OAM and do not consider SOC, the spin degree of freedom can be ignored.
Including spin would merely double the size of the basis, without any coupling between the spin up and down block, making computations more costly.
Hence, we do not consider the spin degree of freedom, while emphasizing that all our results straightforwardly extend to spinful electrons.
Under the assumption of zero SOC, such an extension would simply introduce a spin degeneracy in the energy bands and a doubling of the size of observables.

We work within the atom-centered approximation, only accounting for the OAM associated with the orbitals located on lattice sites \cite{Atencia2024, Johansson2024}, neglecting interatomic or itinerant contributions.
This is well-justified, for example, in elemental ferromagnets such as Ni and Co \cite{Hanke2016}.
In both compounds, $t_{2g}$ orbitals play a key role.
The associated dimensionless OAM operators $\hat{L}_l$, $l=x,y,z$, can be expressed as matrices $\tilde{L}_l$
\begin{equation}
    \tilde{L}_x = \begin{pmatrix}
        0 & 0 & 0 \\
        0 & 0 & i \\
        0 & -i & 0
    \end{pmatrix} \; \tilde{L}_y = \begin{pmatrix}
        0 & 0 & -i \\
        0 & 0 & 0 \\
        i & 0 & 0
    \end{pmatrix} \; \tilde{L}_z = \begin{pmatrix}
        0 & i & 0 \\
        -i & 0 & 0 \\
        0 & 0 & 0
    \end{pmatrix}, \label{L_matrices}
\end{equation}
which act on $\tilde{c}_i$ and $\tilde{c}^\dagger_i$.
We will use the tilde notation whenever the specific matrix representation of the OAM operators in Eq.\ \eqref{L_matrices} is concerned, using $\hat{L}_l$ or $L_l$ otherwise.
$\tilde{L}_l$ satisfy a modified angular momentum algebra, $[\tilde{L}_i, \tilde{L}_j] = -i\varepsilon_{ijl}\tilde{L}_l$, where the additional minus sign arises from truncating the full $d$-orbital space \cite{Stamokostas2018}.
Each $\tilde{L}_l$ has the three eigenvalues: $1, 0, -1$, corresponding to three eigenvectors that depend on $l$.
For concreteness, consider $\tilde{L}_x$. A pure $d_{yz}$ state is an eigenvector of $\tilde{L}_x$ with eigenvalue 0.
The other two eigenvalues of $\tilde{L}_x$ are $\pm 1$, with corresponding eigenvectors $|\psi_{\pm}\rangle_x = d_{xz}\mp i d_{xy}$.
Thus, it follows that for a state to have nonzero OAM, it must be in the form $(d_{xz}+ \mathrm{e}^{i\phi} d_{xy})/\sqrt{2}$, which has $\langle L_x\rangle = -\sin\phi$.
In other words, a state must be a superposition of orbital states with a relative complex phase factor, breaking time-reversal symmetry, to carry a nonzero OAM.
$\tilde{L}_y$ and $\tilde{L}_z$ have similar structure for the eigenvectors $|\psi_{\pm}\rangle_{y/z}$ and $\langle L_{y/z}\rangle$. 
Here we used notation from previous works \cite{Mercaldo2022, Aase2024}.

The Hamiltonian we use was considered in Ref.\ \cite{Mercaldo2022} to study orbital vortices in superconductors. Orbital-preserving hopping processes are made consistent with the point group symmetry of the square lattice $C_{4v}$, where we denote the associated matrix element as $t_{ij}^{\alpha\beta}$.
Similar to Refs.\ \cite{Mercaldo2022, Mercaldo2023}, we also include
an orbital hybridization term $t_m$, coupling $d_{yz}$ to $d_{xy}$ in the $y$ direction and $d_{xz}$ to $d_{xy}$ in the $x$ direction: $t_{i, i\pm e_y}^{yz,xy}=t_{i, i\pm e_x}^{xz,xy} \equiv t_m$, which breaks multiple symmetries \cite{Mercaldo2022}.
They break all mirror symmetries and thus also inversion symmetry, as well as fourfold symmetry.
Out-of-plane symmetry can be broken by interfacing the two-dimensional layer with a substrate and/or vacuum.
The breaking of the remaining symmetries is more complicated as $t_m$ is not consistent with the point group symmetry of the square lattice.
The lattice must thus be distorted for $t_m$ to be nonzero.

However, we still keep to modeling the system as an effective square lattice.
Importantly, we do so under the assumption that $t_m$ arises from an underlying crystal structure, with lower point-group symmetry than the square lattice.
A nonzero $t_m$ is then permissible due to, for example, bonding between the square lattice sites and ligand atoms or by introducing inhomogeneous strain \cite{Mercaldo2023a}, breaking the aforementioned symmetries.
The breaking of these symmetries also affects the $d$ orbitals directly, but as we work within the tight-binding approximation, we will not consider this in detail. 
Instead, we let $t_m$ capture the effects from the underlying symmetry-breaking crystal field.
We note that including a term ($t_m$) which breaks symmetries not present in the lattice itself (square) is a standard treatment in the literature, for instance in square lattices with Rashba spin-orbit interactions.
Two-dimensional noncentrosymmetric oxide heterostructures grown along the [111] crystallographic orientation have been suggested as candidate systems with the appropriate crystal symmetry to host $t_m\neq0$ \cite{Mercaldo2023, Mercaldo2023a}.

\begin{figure}
    \centering
    \includegraphics[width=0.99\linewidth]{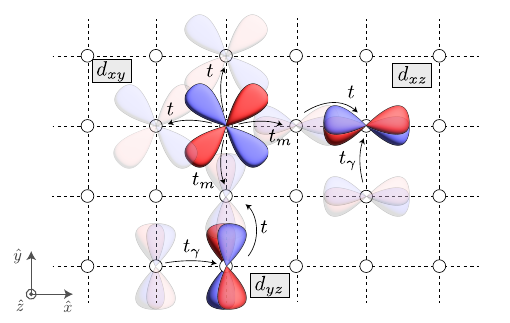}
    \caption{Depiction of the hopping processes in our model. Due to limited wave function overlap, not all hopping between $d$ orbitals is allowed, only those showed here. The figure is adapted and extended from previous work \ \cite{Aase2024}.
    }
    \label{fig:hopping_illustration}
\end{figure}

An illustration of the hopping processes presented here is given in Fig. \ref{fig:hopping_illustration}.
As several symmetries must be broken for $t_m \neq 0$, it is worth dwelling on the relative sizes of hopping magnitudes and other hopping processes allowed by symmetry.
The orbital-preserving processes are kept consistent with the square lattice point group symmetry, as $d_{yz}$ and $d_{xz}$ hoppings are anisotropic $t_{i, i\pm e_x}^{xz,xz} = t_{i, i\pm e_y}^{yz,yz} \equiv t$ and $t_{i, i\pm e_x}^{yz,yz} = t_{i, i\pm e_y}^{xz,xz} \equiv t_\gamma$ and the hopping of $d_{xy}$ orbitals is isotropic $t_{i, i\pm e_x}^{xy,xy} = t_{i, i\pm e_y}^{xy,xy} \equiv t'$.
Because the magnitude of the hopping is determined by the spatial overlap of adjacent atomic orbital wave functions, we take $t'=t$ (which we use as the base energy unit throughout the paper) and consider $t_\gamma<t$.
The difference in spatial overlaps is also the reason why the $t_m$ process is included, while there is no direct coupling between $d_{yz}$ and $d_{xz}$:
the overlap between $d_{xy}$ and the $d_{yz}$ and $d_{xz}$ orbitals is simply larger than that of $d_{yz}$ and $d_{xz}$ since the latter do not have overlapping lobes.
Within our model, $d_{yz}$ and $d_{xz}$ still couple to second order in $t_m$, mediated by their coupling to $d_{xy}$.

Finally, with the breaking of symmetries required for $t_m\neq 0$, the orbital Rashba coupling $t_{\text{SO}}$ can also be nonzero.
In orbitronics, $t_{\text{SO}}$ is known to generate both orbital Hall and Edelstein responses \cite{Go2017}.
In our particular model, the coupling takes the form $t_{\text{SO}}(\sin k_y \tilde{L}_x - \sin k_x \tilde{L}_y)$ in momentum space \cite{Bours2020}.
While symmetry thus allows $t_{\text{SO}}\neq0$ whenever $t_m\neq 0$, because its associated effects are well-studied, we will mainly focus on the physics induced by $t_m\neq 0$, putting $t_{\text{SO}}$ to zero in most of our results.
Still, to connect our framework to previous findings, we will also include some results with $t_{\text{SO}}\neq 0$ when explicitly stating so.

Defining the hopping matrix $\tilde{t}_{ij}$, and only including nearest neighbor hopping, the Hamiltonian is compactly expressed as
\begin{equation}
    H = \sum_{\langle i, j\rangle} \tilde{c_i}^\dagger \tilde{t}_{ij} \tilde{c}_j.
    \label{H_real_space}
\end{equation}
The chemical potential $\mu$ is easily incorporated into $H$ by adding $-\mu \tilde{c}_{i}^\dagger \tilde{c}_{i}$ to Eq.\ \eqref{H_real_space}.
Introducing Fourier-transformed operators $c_{\boldsymbol{k}} = 1/\sqrt{N} \sum_{i}c_{i} \mathrm{e}^{-i\boldsymbol{k}\cdot \boldsymbol{r}_i}$, we rewrite $H$ in momentum space
\begin{widetext}
\begin{align}
    H = -2\sum_{\boldsymbol{k}} \tilde{c}_{\k}^\dagger \begin{pmatrix}
        t_\gamma \cos(k_x)+ t\cos(k_y)+\mu/2 & 0 & t_m\cos(k_y) \\
        0 & t\cos(k_x) + t_\gamma \cos(k_y) +\mu/2 & t_m\cos(k_x) \\
        t_m\cos(k_y) & t_m\cos(k_x) & t(\cos(k_x) + \cos(k_y))+\mu/2
    \end{pmatrix} \tilde{c}_{\k}.
    \label{H_momentum}
\end{align}
\end{widetext}
The matrix in Eq.\ \eqref{H_momentum} is denoted as $\tilde{H}(\k)$.

\begin{figure}
    \centering
    \includegraphics[width=\linewidth]{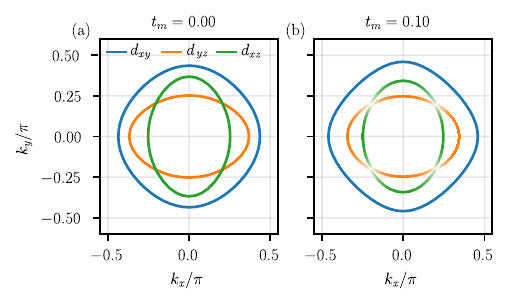}
    \caption{The Fermi surface for two similar systems described by $H(\k)$ in Eq.\ \eqref{H_momentum}. In both systems $t_\gamma=0.5$, $\mu=-2.4$. $t_m$ varies across the panels. The colors indicate which orbital that are the dominant contributor to the energy band, with panel (b) illustrating avoided crossings at $k_x=\pm k_y$, resulting in an interchange of orbital content.
    }
    \label{fig:double_FS}
\end{figure}

For $t_m=0$, Eq.\ \eqref{H_momentum} is already diagonalized; the different $d$ orbitals are eigenvectors corresponding to eigenvalues that can be read off from the diagonal of the matrix in Eq.\ \eqref{H_momentum}. The bandwidth of $\varepsilon_{yz}$ and $\varepsilon_{xz}$ is $4(t+t_\gamma)$, and for $\varepsilon_{xy}$ it is $8t$. Moreover, in the low-filling limit, it follows immediately that the Fermi surface of $d_{xy}$ is a circle, while for $d_{yz}$ and $d_{xz}$ they are perpendicularly orientated ellipses, with the same eccentricity $\sqrt{1-(t_\gamma/t)^2}$. The Fermi surface at intermediate filling is illustrated in Fig.\ \ref{fig:double_FS} (a) and reflects the low-filling behavior.

The Fermi contours corresponding to $d_{yz}$ and $d_{xz}$ orbitals in Fig.\ \ref{fig:double_FS} (a) closely resemble those of the recently discovered altermagnets \cite{Smejkal2022, Smejkal2022a, Noda2016, Hayami2019, Yuan2020, smejkal_sciadv_20, Reimers2024}. 
Microscopic models for altermagnets have been proposed \cite{Brekke2023,Agterberg_2024,Thomale_2024,Franz_2024}, and a range of  transport phenomena in different setups have also been predicted \cite{Gonzalez-Hernandez2021, cui_prb_23,Ouassou2023, attias_prb_24}, thus forming an intriguing backdrop for studying the particular model in Eq.\ \eqref{H_momentum}.
However, effects unique to multiorbital systems—arising from the fact that the $L_l$ operators do not form the full covering group of the angular momentum algebra \cite{Han2022, Han2025}—may alter the transport properties one expects from the altermagnetic studies.

\begin{figure}
    \centering
    \includegraphics[width=\linewidth]{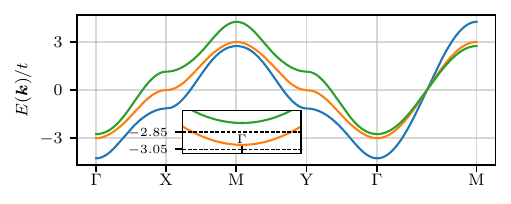}
    \caption{The energy bands for $H(\k)$ in Eq.\ \eqref{H_momentum} with $t_m=0.2$ and $t_\gamma=0.5$ along high-symmetry points. \textit{Inset:} Showing the small $\k$ behavior near $\Gamma$ with dashed lines showing the two values for $\mu$ used in Fig.\ \ref{fig:effect_of_filling_band}.}
    \label{fig:bands}
\end{figure}

With hybridization, $t_m\neq 0$, $\tilde{H}(\k)$ must be diagonalized for all $\k$. 
The Fermi surfaces of these  bands are shown in Fig.\ \ref{fig:double_FS} (b), with otherwise equal parameters as in Fig.\ \ref{fig:double_FS} (a).
We label the three eigenvalues of $\tilde{H}(\k)$ by $\varepsilon_n(\k)$, and in Fig.\ \ref{fig:double_FS} (b) $\varepsilon_1(\k)$ consists of predominantly $d_{xy}$ orbitals, $\varepsilon_2(\k)$ and $\varepsilon_3(\k)$ are the contours with $d_{yz}$ and $d_{xz}$ character at $k_y=0$, respectively.
For $|k_x|=|k_y|$, we see the avoided crossings between bands $\varepsilon_2(\k)$ and $\varepsilon_3(\k)$ induced by $t_m$.
As we illustrate, the corresponding eigenvectors $\psi_1(\k)$ and $\psi_2(\k)$ interchange their orbital character, going from predominantly $d_{yz}$ to $d_{xz}$ character and vice versa.
Both for the sake of completeness and later relevance, the band structure for a system with higher $t_m=0.2$ is given in Fig.\ \ref{fig:bands}.

A seminal paper \cite{Go2018} in orbitronics introduced orbital texture, showing that while $\boldsymbol{L}$ is quenched in equilibrium, an electric field $\boldsymbol{E}$ can induce nonzero $\boldsymbol{L}$, also in centrosymmetric systems. In particular, they found a nonzero orbital Hall conductivity (OHC), $\sigma_{xy}^z$. $\sigma_{ij}^l$ is defined from
\begin{equation}
    J^l_i = \sigma_{ij}^l E_j,
    \label{OHC_def}
\end{equation}
where $J^l_i$ is an OAM current moving in the $i$ direction with orbital polarization in the $l$ direction.
We will shortly see how the orbital texture induces orbital effects in our system.

It is first instructive to compare our model to the paradigmatic model introduced in Ref.\ \cite{Go2018}.
There, the two bands contributing to nonzero $L_z$ have radial and tangential $p$-orbital character, resulting in nonzero $\sigma_{xy}^z$.
These bands consist of $p_x$ and $p_y$ orbitals.
The coupling between $p_x$ and $p_y$ occurs through a second order processes, mediated by an $s$ orbital.
Because Ref.\ \cite{Go2018} has inversion symmetry, it imposes that hopping between $p_{x/y}$ and $s$ must be odd in $k_{x/y}$, resulting in an effective coupling between $p_x$ and $p_y$ that is odd in both $k_x$ and $k_y$.
With the $d$ orbitals we employ, the situation is similar for $L_z$, albeit without inversion symmetry; two orbitals, $d_{yz}$ and $d_{xz}$ are needed to have nonzero $L_z$, and they are only coupled by a second-order process to another orbital, $d_{xy}$.
The difference, however, is that the coupling between $d_{x/yz}$ to $d_{xy}$ is even in $k_{x/y}$, rendering the effective coupling between $d_{yz}$ and $d_{xz}$ even in both $k_x$ and $k_y$.
As we will now see, this difference in the coupling of orbitals leads to a different OAM response to a charge current than the OHE observed in the $sp$-model of Ref.\ \cite{Go2018}.

To showcase the effect of this even coupling analytically, we first consider the case of isotropic diagonal hopping $t_\gamma =t$.
Referring to App.\ \ref{app:pert} for more details, we show that applying an electric field in the $x$ direction induces nonzero $\langle L_l(\k)\rangle$, thus circumventing orbital quenching. 
All bands contribute to $\langle L_l(\k)\rangle$ for $l=x,y,z$.
We calculate the OAM response from the first band explicitly, see Eqs. \eqref{polar_def} and \eqref{L_exp_pert}, and here only highlight some essential features.
The magnitude of the induced $\langle L_l\rangle$ depends on both $|k_x|$ and $|k_y|$.
The sign of $\langle L_l \rangle$, however, is independent of $k_y$ and is the same as the sign of $k_x$.
Following the argumentation of Ref.\ \cite{Go2018}, this is indicative of a nonzero longitudinal orbital conductivity as states with $k_x>0$ and $k_x<0$ will carry positive and negative OAM, respectively.
This also suggests that the orbital Hall conductivity is zero because the induced $L_z$ does not depend on the sign of $k_y$.
The reverse conclusion applies if the effective coupling is in the form $k_xk_y$, consistent with the results and model in Ref.\ \cite{Go2018}.   
Finally, as the magnitude of the induced OAM depends on the magnitude of $k_x$ and $k_y$, there may be both transverse and longitudinal oscillations in $\langle L_l\rangle$ relative the current direction, which will manifest in its real-space profile.
Still, since the sign of $\langle L_l\rangle$ is tied to the sign of $k_x$, it must be odd in the direction of the current and even in the transverse direction, despite the oscillations it may exhibit.

While the calculation in App.\ \ref{app:pert} provides valuable insight into the underlying physics, it has some severe limitations. Specifically, it is only valid for short time scales and can not be used to consider the steady-state of the system.
Perhaps more importantly, considering anisotropic hopping $t_\gamma\neq t$ precludes a tractable analytical diagonalization of Eq.\ \eqref{H_momentum}, making it difficult to extract useful information for such systems.
These  systems are particularly interesting since their anisotropic Fermi surfaces, illustrated in Fig.\ \ref{fig:double_FS}, should manifest in terms of anistropic transport properties.
Additionally, because the orbital character of the bands is momentum-dependent, non-equilibrium-induced OAM effects should become even more pronounced in the presence of anisotropy.
For example, $\varepsilon_2(\k)$ in Fig.\ \ref{fig:double_FS} (b) exhibits a similar radial character as the band considered in Ref.\ \cite{Go2018}, transitioning from predominantly $d_{yz}$ character near the $x$-axis to $d_{xz}$ character near the $y$-axis.
As a last point of note, $t_\gamma<t$ is a requirement for studying systems exhibiting similar Fermi surfaces as spin-up and spin-down electrons in effective models of altermagnets, as mentioned earlier.

So, to extend the simple considerations presented above, we turn to non-equilibrium calculations in the next section. Specifically, we employ the non-equilibrium Green's function technique (NEGF) \cite{datta1997electronic, DATTA2000253} to analyze the steady-state behavior of the system under an applied bias, enabling us to fully capture the interplay between orbital angular momentum and charge currents. 

\section{Non-equilibrium Green function formalism}\label{sec:non_eq}
To drive our multiorbital tight-binding model out of equilibrium, we consider a typical conductor-lead setup where a central conductor region is connected to several leads, each of which is assumed to be in thermal equilibrium with a well-defined chemical potential and temperature, and our approach is inspired by Ref. \cite{nicolic}. The conductor region of finite dimension $N_x \times N_y$ is described by the Hamiltonian as it stands in Eq. (\ref{H_real_space}) while the leads are assumed to be metallic with no orbital character. Observables arising in the central conductor lattice can be obtained in the non-equilibrium formalism through the lesser Green's function $G_{i,j}^<(\omega)$ which is related to the equal-time lattice observable through
\begin{align}
    \langle \tilde{c}_i^{\dagger}\tilde{c}_j \rangle &=-i\lim_{t'\rightarrow t^+}G_{i,j}^<(t-t')\\&=\frac{1}{2\pi i}\int_{-\infty}^\infty \mathrm{d}\omega G_{i,j}^<(\omega)
\end{align}
where $G_{i,j}^<$ is a $3\times3$ matrix in the orbital basis.  

\begin{figure}
    \centering
    \includegraphics[width=0.99\linewidth]{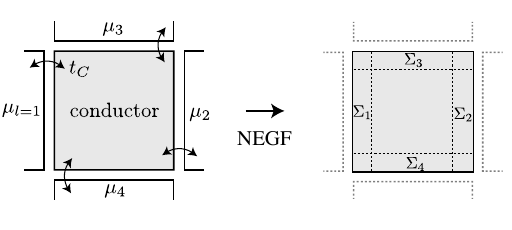}
    \caption{The system is driven out of equilibrium by connecting a multiorbital tight-binding conductor region to a set of infinite leads through a coupling $t_C$. The leads are in thermal equilibrium with a well-defined chemical potential and temperature. We regain a finite system by incorporating the effect of the leads through appropriate self-energy terms on the sites in the conductor which are coupled to the leads by $t_C$. All information about the leads are then stored in these self-energies which can be analytically calculated. Note that at corner sites, self-energy terms from two leads enter. }
    \label{fig: NEGF}
\end{figure}

To obtain the lesser Green's function for the central region biased by the leads, we consider the retarded Green's function operator for the total system consisting of both central region and leads,
\begin{equation}
    G^{R}=\big[(\omega+i\eta)\mathbb{I} - H_\text{tot} \big]^{-1} .\label{eqn: G_r before work}
\end{equation}
As the leads are taken to be infinitely large, the matrix inversion is intractable as it stands, due to the infinite dimensionality of $H_\text{tot}$. Our goal in the following will be to obtain a finite dimensional Green's function for the central region where the infinite dimensions of the leads are taken into account through appropriate self-energies. In the following, both Green's functions and self-energies are functions of energy $\omega$, but the explicit dependence is omitted for brevity of notation.

We proceed by partitioning the Green's function in Eq. (\ref{eqn: G_r before work}) into four sub-matrices, 
\begin{align}
G^{R}&=\begin{bmatrix}
    G_{c}^{R} & G_{cl}^{R} \\
    G_{lc}^{R} & G_{l}^{R}
\end{bmatrix}    \\&= \begin{bmatrix}
    (\omega+i\eta)\mathbb{I} - H_{c} & -H_{{t}}  \\
  -H_{t}^\dagger & (\omega+i\eta)\mathbb{I} - H_{l}
\end{bmatrix}^{-1}    
\end{align}
where $H_{c}$, $H_{l}$ and $H_{t}$ are the Hamiltonians  governing the central region, the leads and the coupling between them respectively. Note in particular that $H_c$ is just the Hamiltonian as introduced in Eq. (\ref{H_real_space}). By multiplying the above expression with $[G^R]^{-1}$ from the left, we can obtain an expression for the finite-dimensional Green's function associated with the center region,
\begin{equation}
    G_\text{c}^{R}=\big[\omega\mathbb{I} - H_\text{c} - \Sigma^R]^{-1}. \label{eqn: retarded Green's function after}
\end{equation}
We have dropped the $i\eta$ due to the presence of $\Sigma^R$ which is sufficient to make the Green function retarded.
The site-representation of the Green function operator is
\begin{align}
    G^R_c(i,j) = \langle i|G^R_c|j\rangle.
\end{align}
This is the quantity that will be used to compute various physical observables. We have introduced the effect of the infinite leads through the self-energy term
\begin{gather}
    \Sigma^R = \sum_{l=1}^4 \Sigma_l^R(i,j) \\ \Sigma_l^R(i,j) =  H_{t}^\dagger G_l^R H_{t}=t_C^2 G_l^R(i_l,j_l) \label{eqn:selfenergyretarded}
\end{gather}
and where we have used that the coupling matrix $H_{t}$ only couples sites on the edge layer of the conductor and sites on the edge layer of the corresponding lead. $G_l^R(i_l,j_l)$ is the Green's function for lead \textit{l} evaluated at the sites $i_l$, $j_l$ on the edge layer in the lead, corresponding to the sites $i$, $j$ on the edge layer in the conductor region. To emphasize, $\Sigma_l^R(i,j)$ will only be non-zero for correlations $(i,j)$ where both $i$ and $j$ are located on the edge layer of the conductor facing the corresponding lead \textit{l}, see Fig. \ref{fig: NEGF}. In the following, we shall denote the central region Green's function defined in Eq. (\ref{eqn: retarded Green's function after}) simply as $G^{R}$. The explicit form of the self-energy term $\Sigma_l(i,j)$ is calculated in App. \ref{app:self-energy}. 

In the Keldysh formalism, $G^<$ is related to $G^R/G^A$ and the self energy through the Langreth rules \cite{langreth},
\begin{equation}
    G^<=G^R \Sigma^< G^A \label{eqn: Keldysh}
\end{equation}
Here we have ignored a transient term focusing solely on the steady-state properties of the system after transients have died out. The expression for the lesser self energy $\Sigma^<$ is obtained by invoking the assumption that each lead is locally in equilibrium with a well-defined chemical potential. As the only operator dependence in the retarded self-energy $\Sigma_l^R$ stems from the retarded lead Green's function $G_l^R$, we can utilize the equilibrium of the leads to obtain
\begin{equation}
    \Sigma^<(\omega)=i\sum_{l=1}^4\Gamma_l(\omega-\mu_l)n_\text{F}(\omega-\mu_l)
\end{equation} 
where $n_\text{F}$ is the Fermi-Dirac distribution and we have defined 
\begin{equation}
    \Gamma_l = i\big(\Sigma_l^R - \Sigma_l^A \big).
\end{equation}
We use an inverse temperature $\beta=100t^{-1}$ throughout this article, corresponding to a temperature of $T\simeq116$ K assuming $t$ to be on the order of 1 eV.
With the lesser Green's function at hand, lattice observables are obtained as
\begin{align}
\langle \hat{O} \rangle =\text{Tr}(\rho \hat{O}) =\frac{1}{2\pi i}\int_{-\infty}^\infty \mathrm{d}\omega \text{ Tr}(\hat{O}  G^<(\omega)) \label{A_obs}
\end{align}

We end this section by noting that in order to take the $t_C\rightarrow 0$ limit, effectively reinstating equilibrium in the conductor region, the self-energy $\Sigma^R$ should be replaced by the usual $i\eta$ factor. Eq. (\ref{eqn: Keldysh}) then reduces to the usual $G^<(\omega)=i A(\omega)n_F(\omega-\mu)$ where $\mu$ is the chemical potential of the conductor region.
We have not considered the equilibrium system in this paper, but have nevertheless included the $i\eta$ factor in addition to the non-zero self-energy contribution from the leads to ensure correct causality. We note, however, that as long as the self-energy contribution from the leads is non-zero, the presence of the $i\eta$ factor does not affect results. $\eta=10^{-10}t$ was used in the calculations. 

\section{Results and discussion}\label{sec:results}
With the non-equilibrium formalism in place, we can now circumvent the ambiguities associated with the OAM current by focusing instead on the local intra-atomic OAM density $\langle L_l\rangle$ emerging under applied bias.
Before moving on to the results, it is worth highlighting why doing so is advantageous.
First, the OAM density is an unambiguous observable and directly measurable by using the MOKE \cite{Choi2023a, Lyalin2023}, unlike the OAM current.
Second, the OAM current is not conserved, so while we will discuss $\sigma_{ij}^l$ from Eq.\ \eqref{OHC_def}, we emphasize that the conductivity of a non-conserved current is conceptually problematic.
Much of the work on the orbital Hall effect relies on calculating $\sigma_{xy}^z$ via the Kubo formula, using it as a figure of merit.
However, since the OAM current is not conserved, it is unclear to what extent a nonzero $\sigma^l$ will give rise to an orbital accumulation.
By instead calculating the OAM density, we are able to directly observe, in real space, the orbital accumulation induced by an electric field.
Moreover, we can compare the actual OAM response with that predicted by computing only the orbital conductivity.
We note in passing that if an orbital Rashba interaction is added to our model, we find that an orbital diode effect appears which is discussed in more detail in App.\ \ref{app:diode}.

\subsection{The orbital splitter effect}

We start by considering the same system as in Fig.\ \ref{fig:double_FS} (b), with a lower chemical potential $\mu=-2.85$.
All energies have been scaled on $t$.
This ensures that all bands are filled, while still being fairly well described by their low-$k$ expansion.
Applying a voltage of $V=0.01$, defined as $\mu_1-\mu_2$, we calculate the lesser Green's functions, in turn allowing calculation of observables through Eq.\ \eqref{A_obs}. We note that a non-zero bias introduces a small non-equilibrium charge modulation in the vicinity of the leads which in a completely rigorous calculation should be calculated self-consistently. We, however, restrict the magnitude of the applied bias to the linear response regime and ignore this correction in the following. This assumption is reasonable as long as the applied bias is low, $eV \ll (E_f - E_b)$ where $E_b$ is the band bottom energy. In effect, the non-equilibrium modulation of the charge density should be small compared to the equilibrium filling level.

\begin{figure}
    \centering   
    \includegraphics[width=\linewidth]{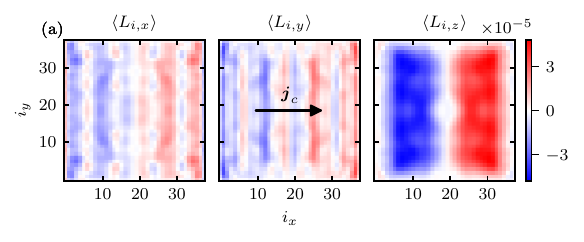}
    
    \vspace{-0.6cm}\hspace{0pt} 

    \includegraphics[width=\linewidth]{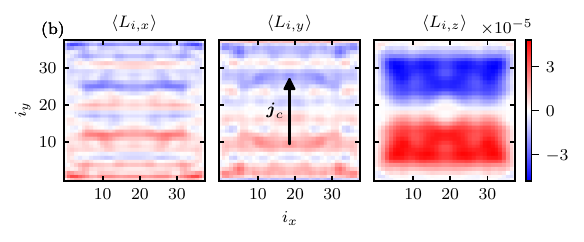}

    \vspace{-0.6cm}\hspace{0pt} 

    \includegraphics[width=\linewidth]{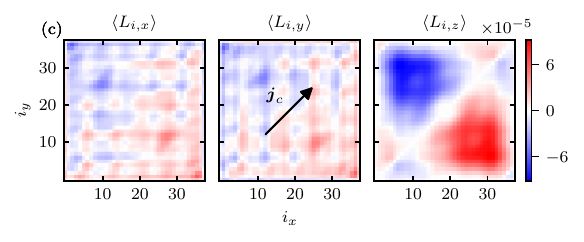}

    \caption{The expectation value for the OAM operators $L_{i,l}$ at each lattice site $i$ for a system with $N \times N$ sites, $N=40$. The system is governed by the Hamiltonian in Eq.\ \eqref{H_real_space} with parameters $\mu=-2.85$, $t_m = 0.1$, $t_\gamma=0.5$, $T=0.01$.
    In all panels, a voltage bias is applied $V = 0.01$ and the black arrows signify the direction of the resulting charge current $\boldsymbol{j}_c$. In (a) and (b), the bias is along one direction: $V_x = V$ in (a) and $V_y=V$ in (b), while in panel (c) biases are applied in both directions simultaneously: $V_x=V_y=V$.}
    \label{fig:longitudinal_orbital}
\end{figure}

In Fig.\ \ref{fig:longitudinal_orbital} (a), the expectation value of the local OAM density is plotted for lattice sites inside the central conductor. 
While this region contains $40\times 40$ sites, we exclude sites that are adjacent to the leads when plotting the results, making the region we plot $N-2\times N-2$.
Let us first consider the symmetries of $\langle L_l\rangle$.
For all $l$, $\langle L_l\rangle$ is odd in the current direction and even in its transverse direction, consistent with the predictions from perturbation theory. 
Thus, the orbital texture stemming from even hybridization gives rise to a OAM accumulation that is consistent with a nonzero longitudinal orbital conductivity. 
We note that in general $L_l$ always vanishes if $t_m=0$, as there is no longer any orbital texture in the system.

The induced OAM density obeys the same symmetries, but $\langle L_l\rangle$ behave differently depending on $l$.
Away from the system edges, $\langle L_z\rangle$ obtains a bulk value, whereas there are both longitudinal and transverse oscillations in $\langle L_x\rangle $ and $\langle L_y\rangle$, relative to the current direction.
These longitudinal oscillations distort the OAM response.
Despite these distortions, for the system under consideration in Fig.\ \ref{fig:longitudinal_orbital} (a), there is still an overall preference to partition $\langle L_{x/y}\rangle$ into a positive and negative region, like $\langle L_z\rangle$.

Figure \ref{fig:longitudinal_orbital} (a) seemingly supports our earlier claim that there is no orbital Hall conductivity in the system.
This is corroborated by results obtained from the Kubo formula for orbital conductivity, which reads, in natural units, \cite{Go2018}
\begin{align}
    \sigma_{ij}^l &= \sum_{n\neq m} \int \frac{\mathrm{d}^2k}{(2\pi)^2} (f_{m\boldsymbol{k}} - f_{n\boldsymbol{k}}) \Omega^{ijl}_{nm\boldsymbol{k}} \label{Kubo_formula} \\
    \Omega^{ijl}_{nm\boldsymbol{k}}  &= \mathrm{Im} \frac{\langle \psi_{n}(\boldsymbol{k})|J_i^l(\k)|\psi_{m}(\boldsymbol{k})\rangle\langle \psi_{m}(\boldsymbol{k})|v_j(\k)|\psi_{n}(\boldsymbol{k})\rangle}{(\varepsilon_{n}(\boldsymbol{k})- \varepsilon_{m}(\boldsymbol{k}) + i\eta)^2},\label{Omega_def}
\end{align}
where $v_j(\k) = \partial_{k_j}H(\k)$ and $J^l_j(\k) = \{v_j(\k), \tilde{L}_l\}/2$. Numerically diagonalizing Eq.\ \eqref{H_momentum} allows straightforward calculation of Eqs.\ \eqref{Kubo_formula} and \eqref{Omega_def}.
As, our work mainly pertains to real-space imaging of the OAM density, we will omit the quantitative results for $\sigma_{ij}^l$ and only state the symmetries it exhibits, relating this to the OAM density. 
In the coordinate system aligned with the crystal axes, we find that for all values of $t_\gamma$ and $t_m$, Eq.\ \eqref{Kubo_formula} yields $\sigma_{ij}^l=0$ when $i\neq j$, consistent with Fig.\ \ref{fig:longitudinal_orbital} (a).
Moreover, the orbital conductivity tensor obeys the following symmetry relations
\begin{align}
    \sigma^z_{xx} = -\sigma^z_{yy} \quad \sigma_{xx}^{x/y} = -\sigma_{yy}^{y/x}.
    \label{sigma_symmetries}
\end{align}
The relative minus signs in Eq.\ \eqref{sigma_symmetries} are non-trivial.
To confirm their existence, in Fig.\ \ref{fig:longitudinal_orbital} (b) we consider instead a nonzero bias in the $y$ direction. The resulting OAM response is consistent with Eq.\ \eqref{sigma_symmetries}; charge current in the positive $y$ direction induces an OAM response with opposite sign relative the current direction compared to Fig.\ \ref{fig:longitudinal_orbital} (a).

As we will now demonstrate, the orbital transverse response is only zero when the charge current is applied in a direction aligning with the crystal axes.
In Fig.\ \ref{fig:longitudinal_orbital} (c), both a vertical and horizontal bias is applied, clearly resulting in a transverse OAM response to the charge current.
Moreover $\langle L_z\rangle$ is zero along the direction of the current.
Both of these points can be understood by examining how the orbital conductivity transforms under rotation around the $z$ axis. In the original coordinate system, the orbital conductivity tensor takes the form $\sigma^l = \text{diag}(\sigma_{xx}^l, \sigma_{yy}^l)$. The transformed conductivity tensor $\sigma'^l$ is then equal to
\begin{align}
    &\sigma'^l = R(\theta)\sigma R^{-1}(\theta) \nonumber \\
    &= \begin{pmatrix}
        \sigma^l_{xx}\cos^2\theta + \sigma^l_{yy}\sin^2\theta & (\sigma^l_{xx}-\sigma^l_{yy})\sin\theta\cos\theta \\
         (\sigma^l_{xx}-\sigma^l_{yy})\sin\theta\cos\theta & \sigma^l_{xx}\sin^2\theta + \sigma^l_{yy}\cos^2\theta
    \end{pmatrix}
    \label{sigma_prime}
\end{align}
where $R(\theta)$ is the conventional 2D rotation matrix with angle $\theta$. Since $\sigma_{xx}^z = -\sigma_{yy}^z$, it follows directly from Eq.\ \eqref{sigma_prime} that in a coordinate system rotated by 45 degrees relative to the crystal axes, only a transverse response remains for the $z$ component of the OAM.
The longitudinal orbital conductivity vanishes, consistent with the behavior we see in Fig.\ \ref{fig:longitudinal_orbital} (c).
However, because $\sigma_{xx}^{x/y}$ and $-\sigma_{yy}^{x/y}$ are not generally equal, a similar cancellation does not occur for $\langle L_x\rangle$ and $\langle L_y\rangle$.
As a result, they exhibit both longitudinal and transverse responses in Fig.\ \ref{fig:longitudinal_orbital} (c), with the transverse response being the dominant one. 

The orbital conductivity closely resemble the spin conductivity in altermagnets.
With suitable alignment with the crystal axis, their longitudinal components in the $x$ direction are opposite to the components in the $y$ direction.
The transverse response to an electric current is therefore the strongest when an electric field is applied in the direction of the band crossings of altermagents, or in our system,  the avoided crossings.
Thus, the effect in Fig.\ \ref{fig:longitudinal_orbital} depends on the orientation relative to the crystal axis, suggesting a comparison to the spin splitter effect in altermagnets \cite{Gonzalez-Hernandez2021, zarzuela_prb_25}. 
The spin-splitter effect has been indirectly observed through torque measurements in RuO$_2$ \cite{Bai2022, Karube2022}.
What we observe here is the orbital counterpart, namely the orbital splitter effect.
States with positive OAM are deflected in one direction, negative OAM in the other, with the effect being dictated by the crystal orientation.
Here, the prerequisite for the orbital splitter effect is $t_m\neq 0$, requiring breaking mirror symmetries and fourfold symmetry.
As mentioned previously, this may be satisfied in oxide heterostructures \cite{Mercaldo2023, Mercaldo2023a}.

Despite the similarities in the splitter effects, the microscopic mechanisms giving rise to the spin and orbital polarizations are different.
The orbital splitter effect presented here arises with the orbital texture, which only exists out of equilibrium, while in altermagnets, the spin polarization, as well as their transport properties, follow directly from their compensated spin bands \cite{Gonzalez-Hernandez2021}, which are polarized even without any current bias.
The equivalent OAM properties are concealed, embedded in the orbital texture, making it far more difficult to obtain a full microscopic explanation of the general OAM behavior.
Moreover, altermagnets host conserved spin currents, in contrast to the nonconserved OAM currents.
Still, the orbital splitter effect is strongly related to the angle between the charge current and the crystal axis, justifying comparison to the spin splitter effect.
Observing it in experiments would require high-quality single crystals to avoid the effect being averaged out.

We note that the distortions in $\langle L_l\rangle$ for the systems in Fig.\ \ref{fig:longitudinal_orbital} are weak enough that the OAM response can be accurately captured using only the orbital conductivity.
As we will see in detail shortly, this is not a generic feature; the induced OAM density may exhibit strong oscillations in both the longitudinal and transverse directions, features that the orbital conductivity is unsuited to describe.

\subsection{OAM oscillations and nonconservation}

\begin{figure}
    \centering
    \stackinset{c}{}{c}{-3.3cm}{\includegraphics[width=\linewidth]{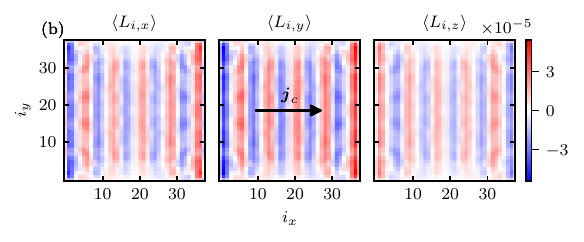}}
    {\includegraphics[width=\linewidth]{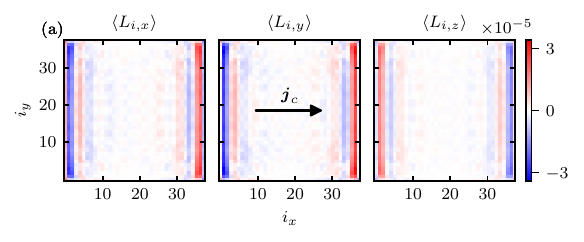}}
    \caption{The same as in Fig.\ \ref{fig:longitudinal_orbital}, with only a horizontal voltage $V_x=V=0.01$. Otherwise the system parameters are $t_m = 0.2$, $t_\gamma=0.5$, $N=40$, $T=0.01$. In panel (a) $\mu=-3.05$ and in panel (b) $\mu=-2.85$. The black arrows signify the direction of the charge current $\boldsymbol{j}_c$.}
    \label{fig:effect_of_filling_band}
\end{figure}

In Fig.\ \ref{fig:longitudinal_orbital}, all bands contribute to the induced OAM.
To disentangle their contributions, we consider two different values for $\mu$, one where only one band is filled, the other where two bands are filled.
We increase $t_m$, in turn increasing the gaps between $\varepsilon_1(\k)$ and $\varepsilon_2(\k)$, as well as between $\varepsilon_2(\k)$ and $\varepsilon_3(\k)$ in Fig.\ \ref{fig:double_FS} (b). 
This allows $\mu$ to be placed between two bands while limiting thermal excitations.
The resulting band structure is shown in Fig.\ \ref{fig:bands}, with the inset showing the dispersion for small $|\k|$.
Applying a horizontal current, we calculate the OAM response in Fig.\ \ref{fig:effect_of_filling_band} for one and two filled bands.
With only one filled band in Fig.\ \ref{fig:effect_of_filling_band} (a), $\langle L_l\rangle$ are oscillating and rapidly diminish away from the leads. 
We observe similar behavior for even lower $\mu$.
However, as we increase $\mu$ to fill the second band, $\langle L_l\rangle$ no longer decays. Their oscillations persist, and the spatial profile is similar for $l=x,y,z$, unlike in Fig.\ \ref{fig:effect_of_filling_band}. 

From Fig. \ref{fig:effect_of_filling_band}, it is evident that the first band plays a minimal role in generating OAM.
This trend persists across various choices for $t_m>0$ and $t_\gamma<t$.
A significant OAM signal emerges only when $\mu$ is large enough to intersect $\varepsilon_2(\k)$. $\varepsilon_3(\k)$ also contributes substantially as it becomes populated.
The simplest explanation for this is that the first band consists of predominantly $d_{xy}$ orbitals, with only minor contributions from $d_{yz}$ and $d_{xz}$. 
Only when the second band is filled does the system begin to rapidly populate more than one orbital, which is necessary for a nonzero OAM, see Eq.\ \eqref{L_matrices}.

\begin{figure}
    \centering
    \includegraphics[width=\linewidth]{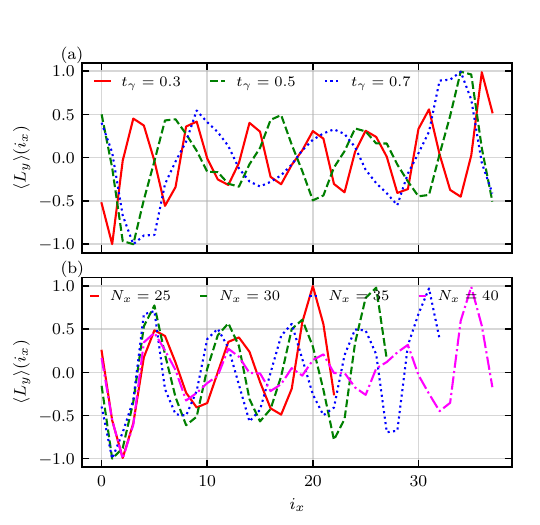}
    \caption{The OAM in the $y$ direction, summed in the vertical direction $\langle L_y\rangle (i_x)\equiv\sum_{i_y} \langle L_{(i_x,i_y), y} \rangle$, as a function of the horizontal position $i_x$.
    To ease comparison, $\langle L_y(i_x)\rangle$ is normalized in each measurement series with respect to its largest value $\max_{i_x} |L_y(i_x)|$
    A horizontal voltage $V_x=V=0.01$ is applied, and otherwise the system parameters are $t_m = 0.25$, $\mu=-2.5$ and $T=0.01$. In panel (a) $N_x=N_y=40$ with $t_\gamma$ varying and in panel (b) $N_x$ is varied while $N_y=40$, $t_\gamma=0.4$ remain constant.}
    \label{fig:oscillations}
\end{figure}

The longitudinal oscillations in Fig.\ \ref{fig:effect_of_filling_band} (b) often feature in systems when two or more bands are filled and the system is far below half-filling. 
Similar oscillations were observed previously in Ref.\ \cite{Han2022} using the quantum Boltzmann formalism, which demonstrated that local injection of OAM results in an oscillatory orbital torsion and OAM as a function of position. 
To examine whether these oscillations are tied to the system size, we vary $N_x$ in Fig.\ \ref{fig:oscillations} (b), keeping all other parameters fixed.
Summing $\langle L_y\rangle$ across rows, we plot the resulting sums as a function of horizontal position.
While there are differences between measurement series, all exhibit periodic variation with approximately the same oscillation length, indicating that the oscillation length depends on microscopic parameters rather than system size.
To explore this dependence further, we vary $t_\gamma$ in Fig.\ \ref{fig:oscillations} (a).
We observe that the oscillation length increases with $t_\gamma$, suggesting that the hopping anisotropy determines the period of the oscillations, as it allows for stronger coupling between the OAM and the lattice. 
This is consistent with previous work \cite{Aase2024}, where we found that decreasing $t_\gamma$ accelerated the absorption of OAM by the lattice.
However, in the present case, the OAM undergoes sustained oscillations rather than being damped, as in Ref.\ \cite{Aase2024}.

\begin{figure}
    \centering
    \includegraphics[width=0.99\linewidth]{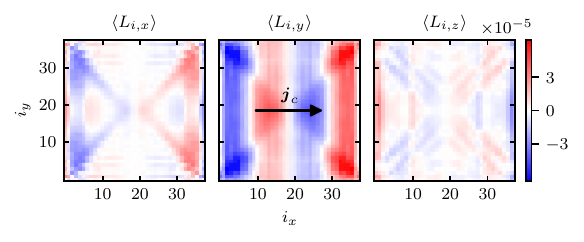}
    \caption{The same as in Fig.\ \ref{fig:longitudinal_orbital}, with different system parameters. A voltage $V_x=V=0.01$ in the $x$ direction is applied, with the black arrow showing the direction of the charge current $\boldsymbol{j}_c$. Otherwise, the system parameters are $N=40$, $t_m=0.1$, $t_\gamma=0.3$, $\mu=-2.0$, and $T=0.1$.}
    \label{fig:higher_filling}
\end{figure}

The oscillations in the OAM densities in Figs.\ \ref{fig:effect_of_filling_band} and \ref{fig:oscillations} could be related to the lack of OAM current conservation.
They showcase the limitations of only relying on $\sigma_{ij}^l$ to describe OAM transport;
because OAM current is not conserved, $\sigma_{ij}^l$ cannot, on its own, accurately describe the OAM behavior in e.g.\ Fig.\ \ref{fig:effect_of_filling_band} (b). 
This is even more obvious when considering a higher filling fraction than previously, with Fig.\ \ref{fig:higher_filling} showing the induced OAM response.
Although some of the effects may be attributed to the Fermi surface being less isotropic at higher filling, it is striking that the OAM response is so intricate, despite the relatively simple system.
We stress that our goal is not to provide a microscopic explanation for Fig.\ \ref{fig:higher_filling}, but rather to demonstrate that some systems defy simple descriptions of the OAM, making lattice frameworks a viable and helpful alternative to solely calculating the orbital conductivity.

\subsection{Orbital Hall effect}

\begin{figure}
    \centering
    \stackinset{c}{}{c}{-3.0cm}{\includegraphics[width=\linewidth]{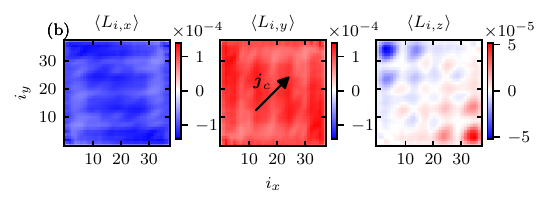}}
    {\includegraphics[width=\linewidth]{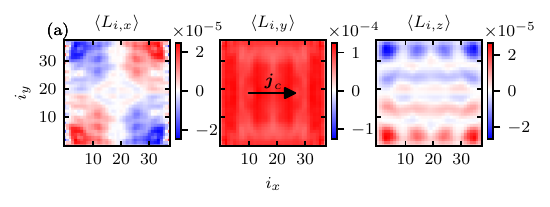}}
    \caption{The same as in Fig.\ \ref{fig:longitudinal_orbital}, with similar system parameters. The differences are that $t_m$ is zero and the orbital Rashba coupling $t_{\text{SO}}=0.1$ is nonzero, otherwise the parameters are the same.
    The black arrows signify the direction of the charge current $\boldsymbol{j}_c$ and varies across the panels.}
    \label{fig:orbital_Rashba}
\end{figure}

As a last demonstration of our framework, we consider a system with $t_m=0$, instead focusing on the orbital Rashba coupling $t_{\text{SO}}$, thus adding $H_{\text{SO}} = t_{\text{SO}}(\sin k_y \tilde{L}_x - \sin k_x \tilde{L}_y)$ to the Hamiltonian in Eq.\ \eqref{H_momentum}.
$H_{\text{SO}}$ is a natural extension from the original $sp$-model \cite{Go2017} to $d$ orbitals \cite{Bours2020}.
The OAM responses to a horizontal and diagonal charge current are shown in Fig.\ \ref{fig:orbital_Rashba} (a) and (b), respectively.
First, the OHE for $\langle L_z\rangle$ can be observed in both panels, showing a genuine transverse OAM response, independent of the current direction, unlike the orbital splitter effect.
The OAM is more localized to the system edges compared to the orbital splitter effect, offering another way to disentangle the two effects.
We hypothesize that the fairly strong distortions of the OAM in the OHE, also seen in the orbital splitter effect, is related to the nonconservation of OAM current.
Second, in both panels, the orbital Edelstein effect can be observed clearly as a bulk value for $\langle \boldsymbol{L}\rangle$ is induced, where the direction of $\boldsymbol{L}$ obeys $\boldsymbol{L}\propto \boldsymbol{z}\times\boldsymbol{j}_c$.
Figure \ref{fig:orbital_Rashba} thus demonstrates that our framework can also reproduce well-known effects, where the spatial resolution may cast them in a new light.

\section{Conclusions}\label{sec:conclusions}
In this work, we have considered non-equilibrium orbital transport and demonstrated the existence of an orbital splitter effect.
Oxide heterostructures may break the required symmetries for this effect to emerge.
OAM accumulation is challenging to predict due to (i) a contribution from multiple bands to the orbital angular momentum accumulation $\langle L_l\rangle$, (ii) $\langle L_l\rangle$ being nonzero generally for $l=x,y,z$, and (iii) the non-conservation of the OAM current.
Our real-space, tight-binding non-equilibrium framework naturally overcomes the former two, and circumvents the latter by enabling direct imaging of the local OAM density across various systems, rather than relying solely on the orbital conductivity associated with the nonconserved OAM current.
The lack of conservation precludes straightforward evaluation of the equation of motions for an OAM operator, as it cannot be written as a continuity equation with the conventional definition of the OAM current.
Instead, torque terms emerge \cite{Go2020a, Han2022, Atencia2024a, Aase2024}, allowing the lattice itself to serve as a source or sink of OAM.

Under applied voltage bias, the OAM response in some cases aligns with the predictions of the orbital conductivity.
However, in most cases we consider, the OAM density varies considerably in real space due to the torque terms, for example oscillating in the direction of the charge current.
Such features cannot be captured by orbital conductivity alone, highlighting the additional insight gained from spatially resolving the OAM.
Our approach of doing so, can, in principle, be readily extended to other systems, by modifying the tight-binding basis and parameters, along with finding suitable representations of the OAM operators. 
For instance, it could be applied to the $sp$ system considered in Ref.\ \cite{Go2018}, or used to investigate the OAM properties of specific materials.

As a final point, we highlight the simplicity of our system.
To obtain a OAM response, we rely solely on applying a voltage bias.
This simplicity is beneficial from an experimental point of view, as it provides a natural way of probing OAM properties.
By coupling a suitable material to leads, making e.g.\ a cross-shaped four-terminal device, one can measure its magnetic response with and without bias.
This is within current experimental capabilities as the orbital Hall effect in titanium was recently measured using MOKE to detect OAM accumulation \cite{Choi2023a}.
For a system with even coupling between orbitals, we predict that the OAM accumulation will depend strongly on which leads that are biased, due to the orbital splitter effect. 

\acknowledgments 
This work was supported by the Research Council of Norway (RCN) through its Centres of Excellence funding scheme, Project No. 262633, "QuSpin", as well as RCN Project No. 323766. 

\begin{widetext}
\appendix

\section{Perturbation theory}\label{app:pert}
In this appendix, we substantiate the claims in the main text regarding the induced OAM in the presence of an electric field. This is done by first-order perturbation theory and follows in spirit the calculation of Ref.\ \cite{Go2018}. As mentioned in the main text, analytically diagonalizing $\tilde{H}(\k)$ from Eq.\ \eqref{H_momentum} with $t_\gamma\neq t$ and $t_m\neq 0$ is possible, but the expressions become too convoluted to extract any information. So, to expedite calculations we consider a system with $t_\gamma = t$. Diagonalizing $H(\k)$ then straightforwardly yields the three eigenvalues
\begin{align}
    \varepsilon_{1/3}(\k) = -2t(\cos k_x + \cos k_y) \mp 2t_m\sqrt{\cos^2k_x + \cos^2k_y)} \quad \varepsilon_2(\k) = -2t(\cos k_x + \cos k_y) \label{lambda_full_k},
\end{align}
and the associated eigenvectors 
\begin{align}
    \psi_{1/3}(\k) = \frac{1}{\sqrt{2}\sqrt{\cos^2k_x + \cos^2k_y}} \begin{pmatrix}
        \cos k_y \\
        \cos k_x \\
        \mp \sqrt{\cos^2k_x + \cos^2k_y}
    \end{pmatrix} \quad \psi_2(\k) = \frac{1}{\sqrt{\cos^2k_x + \cos^2k_y}} \begin{pmatrix}
        \cos k_x \\
        -\cos k_y \\
        0
    \end{pmatrix}.
\end{align}
To proceed in a convenient manner, we do a low $\k$ expansion, neglecting terms $\mathcal{O}(k_{x/y}^4)$. We stress that this is not a necessary simplification for the end result.
The eigenvalues in Eq.\ \eqref{lambda_full_k} then become
\begin{align}
    \varepsilon_{1/3}(\k) = -4t \mp 2\sqrt{2}t_m + \left[ (t \pm t_m/\sqrt{2}) (k_x^2 + k_y^2) \right], \quad \varepsilon_2(\k) = -4t +t(k_x^2+k_y^2), \label{eigenvalues_cont}
\end{align}
resulting in circular Fermi surfaces where the bands have different radii. The eigenvectors similarly simplify
\begin{align}
    \psi_{1/3}(\k) = \frac{1}{2} \begin{pmatrix}
        1+ k_x^2/4 - k_y^2/4 \\
        1 + k_y^2/4 - k_x^2/4 \\
        \mp \sqrt{2}
    \end{pmatrix} \quad
    \psi_2(\k) = \frac{1}{\sqrt{2}} \begin{pmatrix}
        1- k_x^2/4 + k_y^2/4 \\
        -1  - k_x^2/4 +k_y^2/4 \\
        0
    \end{pmatrix}. \label{eigenvector_cont}
\end{align}

Now, we perturb the system by applying a small electric field in the $x$ direction, thus adding the term $e E_x x$ to the Hamiltonian.
In the momentum representation, this is equivalent to $ieE_x \partial_{k_x}$.
Adding this field at $t=0$, the time-evolution of the lowest-energy state $\psi_1(\k)$ at short time scales $\delta t$ is then
\begin{align}
    \ket{\psi_1(\boldsymbol{k}, \delta t)} \approx \mathrm{e}^{-i\varepsilon_1(\k)\delta t} \left[\ket{\psi_1(\boldsymbol{k}, 0)} + ieE_x \sum_{m\neq 1} \frac{\bra{\psi_m(\k)}\partial_{k_x}\ket{\psi_1(\k)}}{\varepsilon_1(\k) - \varepsilon_m(\k)}\left( 1-\mathrm{e}^{i(\varepsilon_1(\k) -\varepsilon_m(\k))\delta t}\right) \ket{\psi_m(\k)}\right].
    \label{pert_theory_expression}
\end{align}
In the interest of brevity, we only include $m=2$ in the above going forward, while noting that the contribution from the third eigenstate in Eq.\ \eqref{pert_theory_expression} will contribute to inducing OAM in a similar way. We return to this later. 
At this stage, we remark that the matrix element $\bra{\psi_2(\k)}\partial_{k_x}\ket{\psi_1(\k)}$ is $\propto k_x$ to lowest order in $k_{x/y}$. If, however, different-parity orbitals are coupled, the lowest-order matrix element would be $\propto k_y$. We comment on the consequences of this shortly.

Inserting Eq.\ \eqref{eigenvector_cont} into Eq.\ \eqref{pert_theory_expression}, keeping only the $m=2$ term, we obtain
\begin{align}
    \ket{\psi_1(\boldsymbol{k}, \delta t)} \approx \mathrm{e}^{-i\varepsilon_1(\k)\delta t}  \left[ \ket{\psi_1(\boldsymbol{k}, 0)} + \frac{ieE_x k_x}{2\sqrt{2}[\varepsilon_1(\k) - \varepsilon_2(\k)]}\left( 1-\mathrm{e}^{i(\varepsilon_1(\k) -\varepsilon_2(\k))\delta t}\right) \ket{\psi_2(\k)}\right].
    \label{pert_theory_simpler}
\end{align}
To continue, we write the prefactor of $\ket{\psi_2(\k)}$ above in its polar form. Both the magnitude $r(\k)$ and the argument $\theta(\k)$ are complicated functions of $\k$, but straightforward to find.
Here, we explicitly extract the factor of $k_x$ from $r(\k)$, along with a factor of $\sqrt{2}$ for convenience, such that we have
\begin{align}
    \frac{ieE_x k_x}{2\sqrt{2}[\varepsilon_1(\k) - \varepsilon_2(\k)]}\left( 1-\mathrm{e}^{i(\varepsilon_1(\k) -\varepsilon_2(\k))\delta t}\right)&= \sqrt{2}k_xr(\k)\mathrm{e}^{i\theta(\k)}.
    \label{polar_def}
\end{align}
Because we extracted $k_x$, it follows directly that both $r(\k)$ and $\theta(\k)$ are even functions of both $k_x$ and $k_y$ since their $\k$ dependence comes from $\varepsilon_1(\k)-\varepsilon_2(\k)$, where only $k_x^2$ and $k_y^2$ enter. This will be important shortly. Otherwise, we note that $r(\k)$ and $\theta(\k)$ vanishes in the limits of $E_x\rightarrow0$ and $\delta t\rightarrow0$, respectively. 

From the definition in Eq.\ \eqref{polar_def}, we insert for $\psi_1(\k)$ and $\psi_2(\k)$ in Eq.\ \eqref{pert_theory_simpler}, such that it reads, in the original basis,
\begin{equation} \label{psi_delta_t}
    \ket{\psi_1(\boldsymbol{k}, \delta t)} = \frac{\mathrm{e}^{-i\varepsilon_1(\k)\delta t}}{2} \begin{pmatrix}
        1 + k_xr(\k)\mathrm{e}^{i\theta(\k)} +  \frac{k_x^2 - k_y^2}{4} (1 - k_xr(\k)\mathrm{e}^{i\theta(\k)}) \\
        1 - k_xr(\k)\mathrm{e}^{i\theta(\k)} - \frac{k_x^2 - k_y^2}{4} (1 + k_xr(\k)\mathrm{e}^{i\theta(\k)}) \\
        -\sqrt{2}
    \end{pmatrix}.
\end{equation}

With Eq.\ \eqref{psi_delta_t}, it is straightforward to find the OAM expectation values for this state by using $\tilde{L}$ from Eq.\ \eqref{L_matrices}. They are
\begin{subequations}
\label{L_exp_pert}
\begin{align}
    \langle L_x \rangle_{\psi_1(\k,\delta t)} &= \frac{k_xr(\k)(4 +k_x^2-k_ y^2)}{4\sqrt{2}}\sin\theta(\k) \\
    \langle L_y \rangle_{\psi_1(\k,\delta t)} &=\frac{k_xr(\k)(4 -k_x^2+k_ y^2)}{4\sqrt{2}}\sin\theta(\k) \\
    \langle L_z \rangle_{\psi_1(\k,\delta t)} &= k_xr(\k)\sin \theta(\k).
    \end{align}
\end{subequations}
From Eqs.\ \eqref{L_exp_pert} three statements made in Sec.\ \ref{sec:orb_texture} immediately follow. First, we observe that the sign of the induced OAM $\langle L_l\rangle$ is the same as the sign of $k_x$. Thus, states with $k_x>0$ carry a nonzero and opposite OAM to the states with $k_x<0$. Since this is induced by a an electric field, also in the horizontal direction, we obtain a longitudinal orbital conductivity, as stated in the main text.
Second, because $\langle L_l \rangle_{\k}$ depend on the magnitudes of $k_x$ and $k_y$, it is reasonable to expect both transverse and longitudinal oscillations of $\langle L_l \rangle$ in real space, which was also stated in the main text.
Third, if the coupling was between orbitals with different parity, as discussed previously, $\theta$ would be proportional to $k_y$, not $k_x$.
The reverse conclusion would apply, and we would then obtain a nonzero orbital Hall conductivity, like in Ref.\ \cite{Go2018}, not a longitudinal one.

The applicability of these statements is, of course, limited to both the perturbative regime and short time scales, as discussed in the main text.
They do not, however, rely on the simplifications used to derive Eq.\ \eqref{L_exp_pert}. 
First, omitting the low $k$ expansion makes for a longer calculation, but the symmetry of the effect remains the same: $k_x$ determines the sign of the induced OAM. 
Second, including transitions from the first to the third band in Eq.\ \eqref{pert_theory_expression} yields similar, terms as in Eq.\ \eqref{psi_delta_t}.
The same is true when considering a system where all bands are filled and considering their respective transitions caused by the electric field, but the sign of $k_x$ still determines the sign of $\langle L_l\rangle$.

\section{Calculation of the lead self energy term \label{app:self-energy}}
We now proceed by calculating the self energy associated with one of the leads connected to the central region. We assume the interface between the central region and the lead to run in the \textit{y} direction. The lead is considered to be of infinite length in the \textit{x} direction.

Assuming the lead to be modeled by a tight-binding model, the transverse eigenfunctions are given by
\begin{equation}
    |\psi_{k_y} \rangle = \sqrt\frac{2}{N_y + 1}\sum_{n_y=1}^{N_y}\sin (k_y n a)|n_y\rangle
\end{equation}
with energy
\begin{equation}
    \varepsilon_{k_y}=-2t_L \cos(k_y a)
\end{equation}
where $t_L$ is the tight-binding hopping parameter in the lead. The longitudinal part of the lead eigenfunctions is given by
\begin{equation}
    |\psi_{k_x}\rangle =\sqrt\frac{2}{N_\text{inf}}\sin(k_x n_x a)|n_x\rangle
\end{equation}
with energy
\begin{equation}
    \varepsilon(k_x)=-2t_L\cos(k_x a)
\end{equation}

We now expand the lead Green's function matrix element $G^R(i_l, j_l)$ in these eigenstates, 
\begin{align}
    \langle i_l | G^R(\omega) | j_l \rangle &= \sum_{k_x, k_y}\frac{\langle i_l|k_x,k_y\rangle\langle k_x,k_y| j_l\rangle}{\omega - 2t_L(\cos(k_xa)+\cos(k_ya)) + i\eta} \\
    &=\sum_{k_y}\langle i_l^y | k_y \rangle \langle k_y |j_l^y \rangle \times \frac{2}{N_\text{inf}}\sum_{k_x}\frac{\sin^2(k_x a)}{\Omega- 2t_L(\cos(k_xa)+\cos(k_ya)) + i\eta}
\end{align}
Utilizing the infinite nature of the lead, we rewrite the $k_x$-sum as an integral, 
\begin{align}
    I(k_y) &= \frac{2}{N_\text{inf}}\sum_{k_x}\frac{\sin^2(k_x a)}{\omega- 2t_L(\cos(k_xa)+\cos(k_ya)) + i\eta} \\
    &= \frac{a}{4\pi t_L} \int_0^{\pi/a}dk_x\frac{2 - e^{2ik_xa} - e^{-2ik_xa}}{(\tilde{\omega} + i\eta)/2t_L- \cos(k_x a)}
\end{align}
where we have defined $\tilde{\omega}=\omega-2t_L\cos(k_y a)$. We convert this into a contour integral over the unit circle,
\begin{align}
    I(k_y) &=\frac{1}{4i\pi t_L}\oint_{|z|=1}dz \frac{1-z^2}{Cz -z^2/2 - 1/2}
\end{align}
where $C=(\tilde{\omega} + i \eta)/2t_L$. The integrand has poles at $z=C \pm \sqrt{C^2 - 1}$. 

If $|\tilde{\omega}| > 2t_L$, the integral becomes 
\begin{equation}
    I(k_y)=\frac{1}{2t_L^2}\bigg(\tilde{\omega} - \text{sgn}(\tilde{\omega})\sqrt{\tilde{\omega}^2 - 4t_L^2} \bigg)
\end{equation}
and likewise for $|\tilde{\omega}| < 2t_L$,
\begin{equation}
    I(k_y)=\frac{1}{2t_L^2}\bigg(\tilde{\omega} - i\sqrt{4t_L^2 - \tilde{\omega}^2} \bigg)
\end{equation}

Finally, we have the expressions for the self-energies for Green's functions describing correlations between sites \textit{i} and \textit{j} where both $(i,j)$ are on the edge layer of the conductor adjacent to the lead. For $|\tilde{\omega}| > 2t_L$,
\begin{equation}
    \Sigma(i,j)=\frac{2}{N_y + 1}\sum_{k_y}\sin(k_y i_y a)\sin(k_y j_y a)\frac{t_c^2}{2t_L^2}\bigg(\tilde{\omega} - \text{sgn}(\tilde{\omega})\sqrt{\tilde{\omega}^2 - 4t_L^2} \bigg)
\end{equation}
while for $|\tilde{\omega}|  < 2t_L$, 
\begin{equation}
    \Sigma(i,j)=\frac{2}{N_y + 1}\sum_{k_y}\sin(k_y i_y a)\sin(k_y j_y a)\frac{t_c^2}{2t_L^2}\bigg(\tilde{\omega} - i\sqrt{4t_L^2 - \tilde{\omega}^2} \bigg)
\end{equation}
\section{Orbital diode effect}\label{app:diode}
\begin{figure}[h!]
    \centering
    \includegraphics[width=0.45\linewidth]{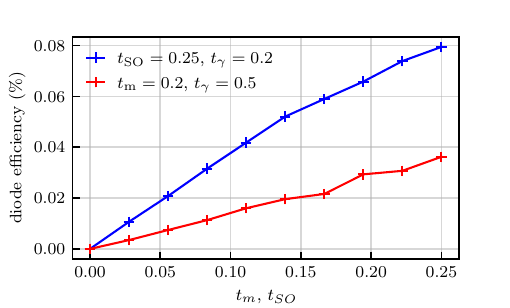}
    \caption{Charge current diode effect is shown as function of $t_m$ for fixed $t_\text{SO}$ (blue) and $t_\text{SO}$ for fixed $t_m$ (red). In this model, one needs both $t_m \neq 0$ and $t_\text{SO}\neq 0$ for a diode effect to manifest in the charge current.  }
    \label{fig: diode effect}
\end{figure}

The diode effect in conventional electronics is a fundamental element in electric circuits, enabling functionality such as rectification. The diode effect consists of the electric current magnitude being different when the polarity of the applied voltage is reversed. In the context of magnetoelectric effects, a charge diode effect is known to occur in the presence of magnetic polarization and spin-orbit interactions. This occurs when the triple product between the current vector $\vecj$, magnetization $\vecM$ and axis of broken inversion symmetry $\vecn$ is non-zero: $\vecj \cdot (\vecM \times \vecn) \neq 0$. 

In the present case, we find a diode effect for the charge current whenever $t_m \neq0$ and orbital Rashba coupling \cite{Park2011} is present.
In momentum space, within our framework, the orbital Rashba coupling takes the form
$H_{\text{SO}} = t_{\text{SO}}(\sin k_y \tilde{L}_x - \sin k_x \tilde{L}_y)$ \cite{Bours2020}, as stated in the main text.
The parameter $t_m$ plays a qualitatively similar role as a spin-dependent mass in kinetic ferromagnets. In the latter case, such a spin-dependent mass provides a spin polarization and net magnetization. In the present case, $t_m$ does not provide any net orbital magnetization on its own, but in conjunction with an applied bias, serves to introduce a non-zero orbital texture.  Together with orbital Rashba coupling, this serves to fulfill the requirements of a charge diode effect.
 
To quantify the charge diode effect, we introduce the diode efficiency and define it as the difference in the cross section current for a given applied voltage difference across the junction $+V$ and the opposite applied voltage difference $-V$, normalized by the forward current. Note that the system used to quantify the diode effect only has two leads, a source and a drain, to ensure conservation of the cross section charge current. The charge diode efficiency is shown in Fig. \ref{fig: diode effect} as function of $t_m$ (blue) and $t_\text{SO}$ (red) and the observed magnitude of the diode effect is in general very small, typically in the $0.01\%-0.1\%$ range. 

Whereas the charge diode effect is well-studied, an orbital diode effect has not been studied previously to the best of our knowledge. Defining an orbital diode effect requires some care. Orbitally polarized currents are in general not conserved quantities, for instance in the presence of orbital Rashba coupling or anisotropic hopping parameters for the different orbitals. This is analogous to how spin currents are generally not conserved in the presence of spin-dependent interactions. Therefore, we instead define an orbital diode effect in terms of a well-defined experimental observable: the accumulation of orbital angular momentum in the system. This plays an equivalent role as spin accumulation in spin transport experiments. We have already established that the presence of a charge current is associated with the emergence of a non-zero OAM polarization, see for instance Fig. \ref{fig:longitudinal_orbital}. Due to the charge diode effect shown in Fig. \ref{fig: diode effect} an OAM polarization is induced in the forward and backward current direction which is not related by a simple sign, but which differs in magnitude. As such, due to the underlying charge diode effect, an asymmetric OAM polarization occurs in our system which we coin the orbital diode effect. Similarly to the charge diode effect in the present system, the orbital diode effect is very small and not useful for practical purposes. However, the concept of an orbital diode effect itself is interesting and may warrant investigation of whether or not a much larger orbital diode effect can occur in other junction setups.

\end{widetext}

\bibliography{main.bib}
\end{document}